\documentclass[11pt]{article}
\usepackage[utf8]{inputenc}
\usepackage[T1]{fontenc}
\usepackage{mathptmx}
\usepackage[a4paper,margin=2.5cm]{geometry}
\usepackage{amsmath,amssymb}
\usepackage{graphicx}
\usepackage{booktabs}
\usepackage{float}
\usepackage{algpseudocode}
\usepackage{cite}
\usepackage{url}
\usepackage[hidelinks]{hyperref}

\title{SAFE: Spatially-Aware Feedback Enhancement for Fault-Tolerant Trust Management in Event-Based VANETs}
\author{İpek Abasıkeleş-Turgut\\[2pt]
\small Computer Engineering Department, Iskenderun Technical University, Hatay, Türkiye\\
\small \texttt{ipek.abasikeles@iste.edu.tr} \quad ORCID: \href{https://orcid.org/0000-0002-5068-969X}{0000-0002-5068-969X}}
\date{}

\begin{document}
\maketitle

\begin{abstract}
In event-based trust management for vehicular ad hoc networks (VANETs), vehicles that witness a road event broadcast its state, and vehicles that later witness the same event score the earlier broadcasters in feedback reports sent to a central decision unit (CDU). When the event state changes, an honest vehicle that reported the previous state receives negative feedback, as the evaluator compares an out-of-date message with the new state. This stale-witness problem is hypothesised to be a major source of unfair penalisation of honest vehicles. To address it, SAFE (Spatially-Aware Feedback Enhancement) is proposed. In SAFE, vehicles continue to record event messages after their decision and throughout the witness area, and send an updated feedback report when they leave it. SAFE was compared with the trust cascading-based emergency message dissemination model (TCEMD) in attack-free highway scenarios simulated with OMNeT++, Veins and Simulation of Urban MObility (SUMO). In the single-event scenario, the negative-feedback rate in the two update rounds after the state change decreased from 53.8\% to 24.7\% and from 55.6\% to 8.3\%, and the number of distinct honest vehicles blacklisted decreased from 20 to 6. In the multi-event scenario, the negative-feedback rate remained at or below 1.3\% in SAFE, compared with up to 77.6\% in TCEMD, and the share of trust evaluations ending in an untrusted label decreased from 22.7\% to at most 1.7\%. These gains required 2.3 to 5.0 times more feedback entries. Experiments with two decision distances showed that a shorter gap between the decision and witness distances reduced stale feedback in both schemes, supporting the stale-witness hypothesis.
\end{abstract}


\section{Introduction}
\label{sec:intro}

Vehicular ad hoc networks (VANETs) are a fundamental component of intelligent transportation systems. They enable vehicles to communicate with each other through vehicle-to-vehicle (V2V) links and with road-side units (RSUs) through vehicle-to-infrastructure (V2I) links. The rapid dissemination of warning messages enables a response to accidents and congestion before they become visible to drivers \cite{Smart}. High mobility and the open wireless medium make it essential to determine which vehicles and messages are trustworthy. Trust-based schemes provide this knowledge with low computational and communication cost, and have consequently been widely adopted in VANETs \cite{CTVAN}.

In event-based trust schemes \cite{DIVA,Dynamic,TCEMD,HTEMD,NOTRINO,RTEAM,DUEL}, a vehicle that witnesses a road event, such as an accident, broadcasts an event message (EM) carrying the event state. Approaching vehicles collect these messages and make a one-time decision at a decision distance, based on their distance to the event and the trustworthiness of the senders. Sender trustworthiness is computed periodically by a central decision unit (CDU) from feedback reports (FBs). When a vehicle witnesses the event itself, it scores each earlier sender as consistent ($+1$) or inconsistent ($-1$) with its own observation and forwards these scores to the CDU through an RSU.

In a network of honest vehicles, all feedback is expected to be positive. Simulations conducted in the preliminary study of this work \cite{previous} revealed that honest vehicles nevertheless receive negative feedback around the moment an event changes state. This inconsistency stems from the timing of the evaluation. A vehicle $v_i$ that passes the event while it is active correctly broadcasts ``active''. A vehicle $v_j$ that arrives after the event has ended observes ``inactive'' and scores $v_i$ as inconsistent, although the message of $v_i$ was correct when it was sent. Such timing-induced errors constitute the \emph{stale-witness problem}. As events change state, these errors accumulate in global trust and can lead to honest vehicles being blacklisted. Distance-based schemes such as the trust cascading-based emergency message dissemination model (TCEMD) \cite{TCEMD} and the hybrid-trust-based emergency message dissemination model (HTEMD) \cite{HTEMD} use distance to determine when and on which evidence a vehicle decides. This study focuses on the next step in the same pipeline: keeping the feedback produced after the decision consistent with the current event state.

This study tests two hypotheses:
\begin{itemize}
    \setlength{\itemsep}{0pt}
    \item[\textbf{H1}] In an event-based trust system, a substantial share of negative feedback, and of the resulting unfair blacklisting, is caused by stale witness information around event state changes.
    \item[\textbf{H2}] If evaluating vehicles keep recording event messages until they leave the witness area and send an updated feedback report, the negative-feedback rate and unfair blacklisting decrease.
\end{itemize}

To test these hypotheses, SAFE (Spatially-Aware Feedback Enhancement) is proposed and compared with TCEMD in single-event, multi-event and decision-distance scenarios. The results confirm both hypotheses. In the single-event scenario, SAFE reduced the negative-feedback rate after the state change from 55.6\% to 8.3\% and the number of blacklisted honest vehicles from 20 to 6. In the multi-event scenario, the negative-feedback rate remained at or below 1.3\% in SAFE, while it reached 77.6\% in TCEMD.

The main contributions of this study are as follows:
\begin{itemize}
    \setlength{\itemsep}{0pt}
    \item The stale-witness problem is defined and characterised as a fault-tolerance issue of event-based trust systems.
    \item SAFE is proposed as a feedback-collection mechanism that can be integrated into distance-based event schemes.
    \item The benefit of SAFE (negative-feedback rate, blacklisting, node classification) and its cost (feedback volume, storage, computation) are quantified. The contribution of each SAFE component is analysed from the simulation logs.
    \item The constraint $D_d - D_w \le R$ is derived, linking the decision distance ($D_d$), the witness distance ($D_w$) and the communication range ($R$). The effect of the decision distance on stale feedback is demonstrated.
\end{itemize}

The rest of the paper is organised as follows. \autoref{sec:related} positions this study within the literature. \autoref{sec:model} describes the system, event and trust models and the SAFE approach. \autoref{sec:setup} presents the simulation setup and metrics, and \autoref{sec:results} reports the results. \autoref{sec:discussion} discusses the findings, and \autoref{sec:conclusion} concludes the paper.

\section{Related Work and Positioning}
\label{sec:related}

Event-based trust schemes for VANETs evaluate warning messages together with the vehicles that send them. They differ mainly in how trust is computed and aggregated. Reputation can be stored on a distributed ledger \cite{DIVA}, updated dynamically \cite{Dynamic} or combined from entity and data trust \cite{NOTRINO}. Other schemes evaluate risk \cite{RTEAM} or handle uncertainty with Dempster--Shafer theory \cite{DUEL}. Among these schemes, TCEMD \cite{TCEMD} introduced a distance-based treatment of emergency messages. A vehicle ignores messages about distant events, records messages about nearby events and makes a single decision when it reaches the decision distance. When it later witnesses the event, it produces feedback on the recorded senders, which the CDU aggregates into global trust. HTEMD \cite{HTEMD} adopted the same distance-based structure in a hybrid trust model. In these schemes, distance governs the decision: it determines when, and on which evidence, a vehicle acts on a warning.

SAFE builds on this distance-based structure and extends it to the feedback produced after the decision. After the decision, the recorded evidence continues to determine the feedback that the vehicle reports. SAFE keeps this evidence up to date with two elements: continued recording after the decision and inside the witness area, and an updated feedback report sent when the vehicle leaves the witness area. These elements are not specified in the event-based schemes examined in this study \cite{DIVA,Dynamic,TCEMD,HTEMD,NOTRINO,RTEAM,DUEL}. Since SAFE operates only on feedback collection, it complements the distance-based decision of TCEMD and HTEMD and can be integrated into either scheme.

The preliminary study of this work \cite{previous} developed the event-based simulation framework used here. The framework includes an event class with type, location, duration and effective distance, and adapted versions of two centralised trust schemes. That study analysed the effect of selfish vehicles that refuse to broadcast event messages or to send feedback reports. During that analysis, honest vehicles were observed to receive negative feedback when events changed state. The present study builds on this observation in three ways: (i) it isolates and characterises these non-malicious errors in an attack-free network; (ii) it proposes SAFE as a mechanism to reduce them; and (iii) it analyses the role of the decision distance.

\section{System Model and the SAFE Approach}
\label{sec:model}

This section introduces the system architecture and assumptions, the event model, the distance zones, the baseline trust model and the proposed SAFE approach.

\subsection{Architecture and assumptions}

\begin{figure*}[!htbp]
\centering
\includegraphics[width=0.85\textwidth]{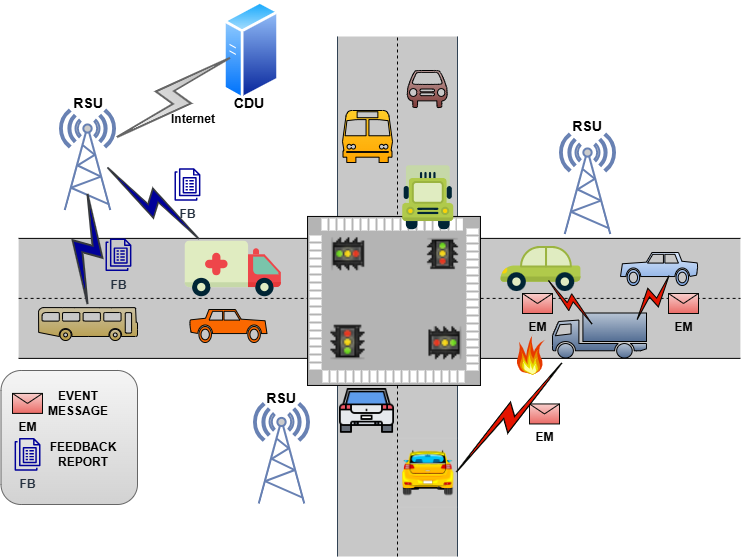}
\caption{Event-based VANET architecture. EM: event message; FB: feedback report. Drawn by the author.}\label{fig:arch}
\end{figure*}

The modelled system (\autoref{fig:arch}) consists of three types of elements: vehicles, RSUs and the CDU. Each vehicle is equipped with an on-board unit (OBU), which comprises an IEEE 802.11p radio and a processing unit. RSUs relay messages between vehicles and the CDU over a wired backbone. The CDU periodically computes the global trust (GT) values of vehicles from the received feedback reports and distributes them to vehicles through RSUs. \autoref{tab:notation} lists the abbreviations and symbols used in this paper.

\begin{table}[!htbp]
\centering
\caption{Abbreviations and notation}
\label{tab:notation}
\begin{tabular}{p{0.18\textwidth}p{0.7\textwidth}}
\toprule
Symbol & Meaning \\
\midrule
OBU & On-board unit (IEEE 802.11p radio and processor) \\
RSU / CDU & Road-side unit / central decision unit \\
GPS & Global positioning system \\
EM / FB & Event message / feedback report \\
TCEMD & Trust cascading-based emergency message dissemination model \cite{TCEMD} \\
GT$_e(i)$ & Global trust of vehicle $i$ computed from feedback on event $e$ \\
$D_w$ & Witness (effective) distance of an event \\
$D_d$ & Decision distance \\
$D_i$ & Distance of interest \\
$D_e$ & Current distance between a vehicle and an event \\
$R$ & Communication range of vehicles \\
$Th_B$, $InTr$ & Blacklist threshold (0.2), initial trust (0.5) \\
FBR & Number of feedback entries received by the CDU in a round \\
UGT & Set (and number) of vehicles whose GT is updated in a round \\
BLR, N\_BLR & Blacklist rate, non-blacklist rate \\
\bottomrule
\end{tabular}
\end{table}

The system assumptions follow the literature:
\begin{itemize}
    \setlength{\itemsep}{0pt}
    \item Vehicles communicate with each other and with RSUs through IEEE 802.11p OBUs \cite{TCEMD,AATMS,RSMA,MDT,HTMS}. Vehicles know their own position and the event position through the global positioning system (GPS) or a similar system \cite{TCEMD,NOTRINO}, and all clocks are synchronised \cite{RSMA}.
    \item Vehicles register with the CDU when they join the network \cite{RSMA}. RSU coverage along the road is sufficient for every feedback report to reach the CDU \cite{AATMS,TCEMD}. The CDU has sufficient storage and computing resources \cite{TCEMD,AATMS,RSMA,MDT}.
    \item RSUs and the CDU are trusted \cite{AATMS,TCEMD,RSMA}. \autoref{sec:security} outlines how this assumption can be relaxed.
    \item All vehicles are honest. Under this assumption, every negative feedback entry is an error and every blacklisted vehicle is a false positive. This setting isolates the fault-tolerance behaviour targeted by SAFE (\autoref{sec:threat}).
\end{itemize}

\subsection{Event model}
\label{sec:event}

An \emph{event} is a road condition that concerns vehicles approaching a given location, such as an accident, a blocked lane or a temporary hazard. Each event has a location, a type, a start time $T_{start}$ and a stop time $T_{stop}$. Its state is 1 (present) between $T_{start}$ and $T_{stop}$ and 0 (absent) otherwise. Events are generated by the simulation. The OBU of a vehicle detects an event automatically when the vehicle enters the effective distance $D_w$ of the event. The OBU then broadcasts an EM containing the event identifier, the observed state, the event position and a time stamp.

The \emph{decision} of a vehicle about an event is binary. The vehicle either accepts the warning and prepares an action, such as slowing down, changing lane or re-routing, or rejects it. The decision is made once, at distance $D_d$, using the messages received so far and the GT values of their senders.

\begin{figure}[!htbp]
\centering
\includegraphics[width=0.55\textwidth]{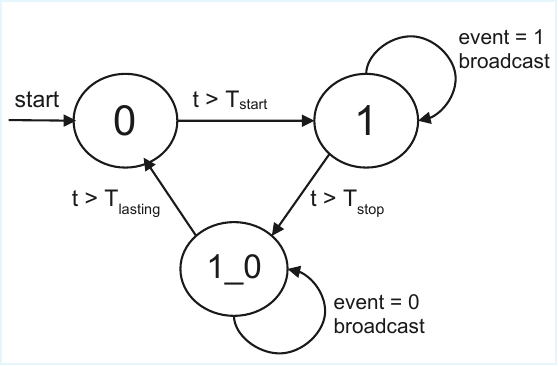}
\caption{Witness broadcast behaviour over the lifetime of an event. In state 1, witnesses broadcast ``present''; in state 1\_0 (between $T_{stop}$ and $T_{lasting}$), they broadcast ``absent''; after $T_{lasting}$, nothing is broadcast.}\label{fig:fsm}
\end{figure}

\autoref{fig:fsm} shows the broadcast behaviour of witnesses over the lifetime of an event. Two end times are defined. At $T_{stop}$, the event disappears and witnesses start broadcasting state 0, so that approaching vehicles learn that the hazard has cleared. Broadcasting state 0 indefinitely would cause every passing vehicle to keep sending EMs about a non-existent event and would lead to a broadcast storm. Therefore, $T_{lasting}$ bounds the period in which state 0 is announced, and no EM is broadcast for the event after $T_{lasting}$. Stale-witness errors arise within the interval $[T_{stop}, T_{lasting}]$.

Events are grouped into three types (\autoref{tab:types}). The effective distances span one order of magnitude, from a local hazard to a highway-scale incident, and thus cover a broad range of event scales. In the simulations, durations are scaled down so that several state changes occur within the simulated time.

\begin{table*}[!htbp]
\centering
\caption{Event types. The durations used in the simulations are given in \autoref{tab:setup} and \autoref{tab:multi}.}
\label{tab:types}
\begin{tabular}{p{0.2\textwidth}p{0.22\textwidth}p{0.22\textwidth}p{0.2\textwidth}}
\toprule
 & Type 1 & Type 2 & Type 3 \\
\midrule
Severity & Low & Medium & High \\
Example & Local hazard, traffic-light warning & Side-road accident, congestion & Main-road accident \\
Effective distance $D_w$ & 100 m & 400 m & 800 m \\
Nominal duration & 10 min & 45 min & 2 h \\
\bottomrule
\end{tabular}
\end{table*}

\subsection{Distance zones}

\begin{figure}[!htbp]
\centering
\includegraphics[width=0.45\textwidth]{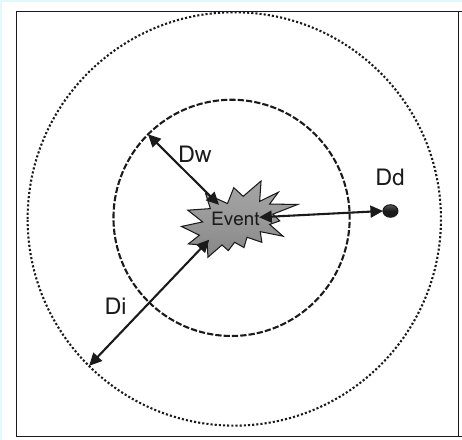}
\caption{Distance zones around an event: witness distance $D_w$, decision distance $D_d$ and distance of interest $D_i$.}\label{fig:zones}
\end{figure}

Following TCEMD \cite{TCEMD}, the behaviour of a vehicle depends on its distance $D_e$ to the event (\autoref{fig:zones}). The zones are nested such that $D_w < D_d < D_i$. A vehicle decides before it reaches the event, which implies $D_d > D_w$, and it ignores events beyond the distance of interest $D_i$. \autoref{tab:zones} lists the actions taken in each zone by TCEMD and SAFE. Intervals are written as $(a, b]$ and denote distance ranges.

\begin{table*}[!htbp]
\centering
\caption{Actions of a vehicle in each distance zone. Differences between the schemes are shown in bold.}
\label{tab:zones}
\begin{tabular}{lll}
\toprule
Zone ($D_e$) & TCEMD & SAFE \\
\midrule
$D_e > D_i$ & Ignore EMs & Ignore EMs \\
$(D_d, D_i]$ & Record EMs & Record EMs \\
$D_e = D_d$ (first time) & Decide once & Decide once \\
$(D_w, D_d)$ & -- & \textbf{Keep recording EMs} \\
$D_e \le D_w$, on entry & Broadcast state, send FB & Broadcast state, send FB \\
$D_e \le D_w$, inside & Broadcast state & Broadcast state, \textbf{keep recording EMs} \\
Leaving $D_w$ & -- & \textbf{Send one updated FB if new evidence exists} \\
\bottomrule
\end{tabular}
\end{table*}

\textbf{Communication constraint.} At the decision point, a vehicle must be able to receive messages directly from the witnesses. Since the nearest witness position on the approaching side lies at distance $D_w$ from the event, the gap between the decision point and the nearest witness is $D_d - D_w$. Single-hop reception at the decision point therefore requires
\begin{equation}
\label{eq:constraint}
D_d - D_w \le R .
\end{equation}
The decision distance itself may exceed $R$. For instance, the Type~3 configuration in \autoref{sec:setup} uses $D_d = 1000$~m with $R = 300$~m. In addition, $D_d - D_w$ must leave enough time for the vehicle to act, which sets a lower bound depending on speed and reaction time. The same gap also determines the age of the evidence in TCEMD: a vehicle that stops recording at $D_d$ holds messages that are at least $(D_d - D_w)/v$ seconds old when it starts witnessing at speed $v$. \autoref{sec:dd} examines this relationship.

\subsection{Baseline trust model (TCEMD)}
\label{sec:tcemd}

SAFE uses the TCEMD trust computation without modification; consequently, the two schemes differ only in the feedback that reaches the CDU. When a vehicle $r$ witnesses event $e$, it compares the latest recorded EM of each sender $i$ with its own observation. It then produces a time-stamped feedback entry $\langle r, i, e, f_{r,i} \rangle$, where $f_{r,i}=+1$ if the states match and $f_{r,i}=-1$ otherwise. The evaluator's own observation serves as the reference, as every vehicle observes the true state within $D_w$.

Every 60~s, the CDU keeps the latest entry for each $\langle r,i,e\rangle$ triple and computes an event-level GT for each target $i$ that received feedback \cite{TCEMD}:
\begin{equation}
\label{eq:gt}
GT_e(i) = \frac{W_b(i) + W_r(i)}{2} \cdot \frac{\sum_{r} f_{r,i}\, GT(r)}{\sum_{r} GT(r)},
\end{equation}
where the sums run over the non-blacklisted reporters of $i$ for event $e$, and $GT(r)$ is the current trust of reporter $r$. With scores in $\{+1,-1\}$, the ratio, and hence $GT_e(i)$, lies in $[-1,1]$. The activity weights reward vehicles that are frequently evaluated and that frequently report:
\begin{equation}
\label{eq:w}
W_x(i) = \begin{cases} 1, & S_x(i) \le |VN|/2 \\ 1.5 - S_x(i)/|VN|, & \text{otherwise} \end{cases} \qquad x \in \{b, r\},
\end{equation}
where $S_b(i)$ and $S_r(i)$ are the ranks of $i$ among the $|VN|$ non-blacklisted vehicles by the number of times it was evaluated and the number of evaluations it made, respectively. As $W_x \in [0.5, 1]$, a vehicle with only positive feedback obtains $GT_e \in [0.5, 1]$. It follows that only negative feedback can lower a vehicle below $InTr=0.5$ and, in particular, to or below $Th_B=0.2$.

After each round, each vehicle is labelled per event as
\begin{equation}
\label{eq:label}
Label_e(i) =
\begin{cases}
\text{Untrusted}, & GT_e(i) \le Th_B \\
\text{Suspicious}, & Th_B < GT_e(i) \le InTr \\
\text{Honest}, & GT_e(i) > InTr.
\end{cases}
\end{equation}
A vehicle labelled untrusted is blacklisted. Once a vehicle is blacklisted, its feedback reports and event messages are ignored in all subsequent rounds and in all events. It may still appear as the target of feedback from other vehicles; in such cases, its GT may continue to be computed.

\subsection{SAFE}
\label{sec:safe}

In TCEMD, a vehicle stops recording at $D_d$ and sends a single FB when it first enters the witness area. Its evidence about each sender is thus limited to the last EM received before the decision point, and this evidence becomes stale if the event changes state in the meantime. SAFE introduces two elements to keep the evidence current (\autoref{tab:zones} and Algorithm~\ref{alg:safe}):
\begin{enumerate}
    \setlength{\itemsep}{0pt}
    \item \emph{Continued recording (C1):} after the decision, the vehicle continues to record EMs in $(D_w, D_d)$ and throughout the witness area. The decision itself remains final, and the new records are used for feedback.
    \item \emph{Updated exit report (C2):} when the vehicle leaves the witness area, it sends one additional FB if it has recorded new evidence since its first FB. At most one additional report is sent per event, which keeps the reporting volume bounded.
\end{enumerate}

\begin{table*}[!htbp]
\centering
\caption{SAFE, vehicle-side procedure for vehicle $v$ and event $e$}
\label{alg:safe}
\begin{tabular}{p{0.95\textwidth}}
\hline
\begin{minipage}{0.93\textwidth}
\vspace{2pt}
\begin{algorithmic}[1]
\State $Rec_e \gets \emptyset$; $decided \gets$ false; $fbSent \gets$ false
\Procedure{OnEM}{$m=\langle j, e, s, t\rangle$}
  \State $D_e \gets \mathrm{dist}(v, \mathrm{loc}(e))$
  \If{$D_e > D_i$ \textbf{or} $j$ is blacklisted} \State \textbf{discard} $m$ \EndIf
  \State $Rec_e[j] \gets (s, t)$ \Comment{keep the latest EM per sender; TCEMD stops once $decided$}
\EndProcedure
\Procedure{OnMove}{}
  \State $D_e \gets \mathrm{dist}(v, \mathrm{loc}(e))$
  \If{$D_e \le D_d$ \textbf{and not} $decided$}
     \State $\mathrm{Decide}(Rec_e, GT)$; $decided \gets$ true \Comment{unchanged decision rule}
  \EndIf
  \If{$D_e \le D_w$}
     \State $\sigma \gets$ observed state of $e$; broadcast $\langle v, e, \sigma, now\rangle$ if $now < T_{lasting}$
     \If{\textbf{not} $fbSent$}
        \State send $\mathrm{FB}(Rec_e, \sigma)$ to RSU; $fbSent \gets$ true; mark $Rec_e$ as reported
     \EndIf
  \ElsIf{$v$ has just left the witness area \textbf{and} $Rec_e$ has unreported entries}
     \State send $\mathrm{FB}(Rec_e, \sigma_{last})$ to RSU \Comment{C2: one updated report}
  \EndIf
\EndProcedure
\Function{FB}{$Rec, \sigma$}
  \State \Return $\{\langle v, j, e, [s_j = \sigma]\rangle : (s_j, t_j) \in Rec\}$
\EndFunction
\end{algorithmic}
\vspace{2pt}
\end{minipage}\\
\hline
\end{tabular}
\end{table*}

Continued recording has two effects. First, the latest record of each sender is more recent, which reduces the number of senders compared with an outdated state. Second, the vehicle receives messages from more senders. As a result, the CDU receives more independent feedback per target, and any remaining stale entry carries less weight in Eq.~(\ref{eq:gt}). \autoref{sec:mech} examines the contribution of each element using the simulation logs.

\subsection{Scope}
\label{sec:threat}

This study focuses on fault tolerance: it measures how often a network of honest vehicles penalises its own members, and how effectively SAFE prevents this. All vehicles report truthfully, and RSUs and the CDU are trusted. In this setting, errors arise from the timing of evaluations and from message losses in the simulated IEEE 802.11p channel; no artificial errors are injected. Any vehicle whose GT decreases necessarily corresponds to an honest vehicle that has been penalised unfairly, which allows the effect of SAFE to be measured directly. \autoref{sec:security} discusses how SAFE can be extended to adversarial settings.

\section{Simulation Setup and Metrics}
\label{sec:setup}

SAFE and TCEMD were implemented in OMNeT++ \cite{omnet++} with Veins \cite{veins} and SUMO \cite{sumo}. Both schemes share the same event model, mobility traces, message formats and trust computation (Eqs.~(\ref{eq:gt})--(\ref{eq:label})), and differ only in the feedback-collection behaviour given in \autoref{tab:zones}. Messages are carried by the Veins IEEE 802.11p stack, and no artificial message error rate is injected. The communication range $R=300$~m is modelled at the application layer: a vehicle treats another node as a neighbour, and processes its messages, only if the Euclidean distance between them is at most $R$.

For a fixed route file and seed, the simulation is deterministic, and a rerun reproduces the same trace. Accordingly, each configuration was simulated once. \autoref{tab:setup} lists the parameters of the single-event scenario.

\begin{table}[!htbp]
\centering
\caption{Simulation parameters (single-event scenario)}
\label{tab:setup}
\begin{tabular}{p{0.42\textwidth}p{0.45\textwidth}}
\toprule
Parameter & Value \\
\midrule
Road & 8000 m, 3-lane, one-way highway \\
Communication range $R$ & 300 m (application layer, Euclidean distance) \\
Event & Type 1 at ($x=2000$ m, $y=400$ m) \\
$T_{start}$ / duration & 50 s / 150 s \\
$T_{stop}$ / $T_{lasting}$ & 200 s / 275 s \\
$D_w$ / $D_d$ / $D_i$ & 100 m / 300 m (also 200 m) / 500 m \\
GT update period & 60 s \\
Initial trust $InTr$ / blacklist threshold $Th_B$ & 0.5 / 0.2 \\
Runs per configuration & 1 (deterministic) \\
\bottomrule
\end{tabular}
\end{table}

\textbf{Parameter selection.} $InTr=0.5$ is a neutral initial value: a newly joined vehicle is neither trusted nor distrusted, and rises above this value only through positive feedback. $Th_B=0.2$ was adopted from the preliminary study of this work \cite{previous} and applied identically to both schemes. $D_w$ follows the event type, and $D_d$ was selected to satisfy Eq.~(\ref{eq:constraint}) with $R=300$~m. Finally, $D_i = D_d + 200$~m provides each vehicle with a recording zone before the decision point.

\textbf{Performance metrics.} Let $\mathcal{F}_k$ denote the multiset of feedback entries received by the CDU in round $k$, $\mathcal{F}_k^-$ its negative entries, and $U_k$ the set of vehicles whose GT was updated in round $k$. The following metrics are used:
\begin{itemize}
    \setlength{\itemsep}{0pt}
    \item Number of feedback entries, $\mathrm{FBR}_k = |\mathcal{F}_k|$. This metric reflects both the evidence available to the CDU and the communication cost.
    \item Number of updated global trusts, $\mathrm{UGT}_k = |U_k|$.
    \item Negative-feedback rate, $|\mathcal{F}_k^-| / |\mathcal{F}_k|$. In a network of honest vehicles, every negative entry is an error. The positive-feedback rate is its complement.
    \item Blacklist rate and non-blacklist rate:
    \begin{align*}
    \mathrm{BLR}_k &= \frac{|\{\, i \in U_k : GT(i) \le Th_B \,\}|}{|U_k|}, \\
    \mathrm{N\_BLR}_k &= 1 - \mathrm{BLR}_k .
    \end{align*}
    $\mathrm{BLR}_k$ is the fraction of updated vehicles that are blacklisted; here, every blacklisting is a false positive.
    \item Distinct blacklisted vehicles: the number of different vehicles blacklisted at least once during a run.
    \item Node classification according to Eq.~(\ref{eq:label}), counted as \emph{node-evaluations}: a vehicle evaluated in several rounds or events is counted once per evaluation.
    \item Rejected warnings: decisions at $D_d$ in which the vehicle judges the received event message to be incorrect and does not act on it. As all vehicles are honest, each rejection constitutes an erroneous decision.
\end{itemize}
GT rounds are identified by their time (60~s, 120~s, \dots), which allows the single-event and multi-event scenarios to be compared on a common time axis.

\section{Results}
\label{sec:results}

This section presents the results of the single-event and multi-event scenarios, analyses the contribution of the SAFE elements and the effect of the decision distance, and evaluates the overhead of SAFE.

\subsection{Single-event scenario}
\label{sec:single}

\begin{figure*}[!htbp]
\centering
\includegraphics[width=\textwidth]{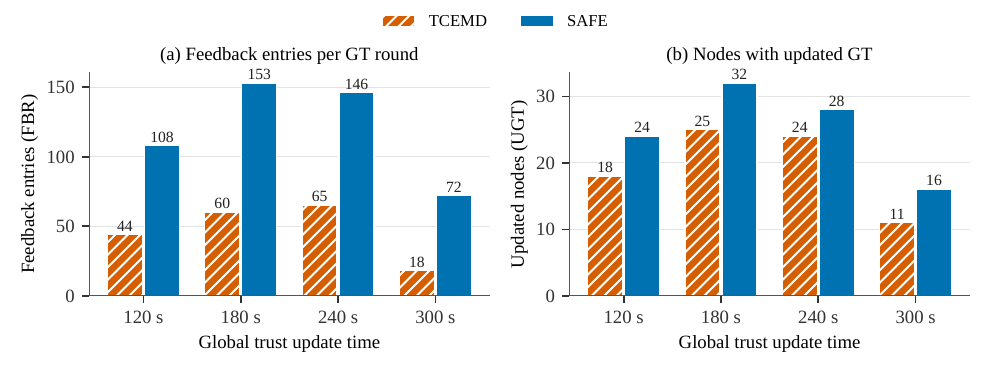}
\caption{Single-event scenario ($D_d=300$~m): (a) feedback entries and (b) vehicles with updated GT in each GT round.}\label{fig:single_fbr}
\end{figure*}

\autoref{fig:single_fbr}(a) shows the number of feedback entries per round. SAFE delivered more feedback entries than TCEMD in every round, with 108 and 44 entries at 120~s, 153 and 60 at 180~s, 146 and 65 at 240~s, and 72 and 18 at 300~s, respectively. Over the entire run, SAFE collected 482 entries against 187, corresponding to a 2.6-fold increase. With this richer evidence, the CDU updated the GT of more vehicles in each round (\autoref{fig:single_fbr}(b)), for example, 32 and 25 vehicles at 180~s, respectively.

\begin{figure*}[!htbp]
\centering
\includegraphics[width=\textwidth]{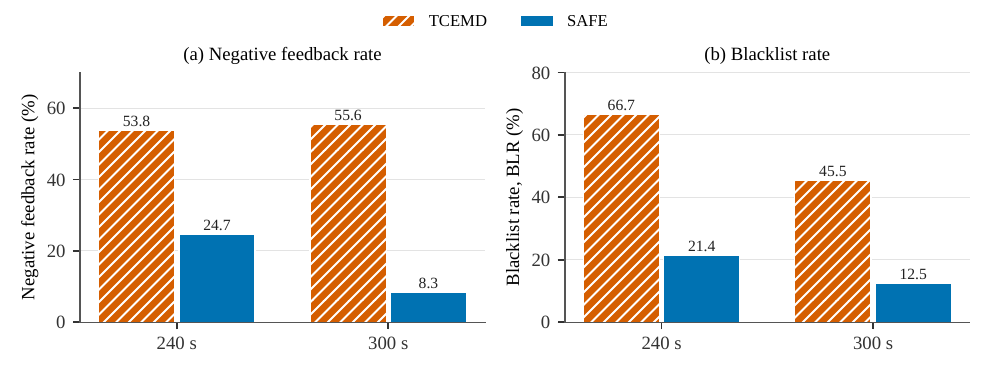}
\caption{Single-event scenario ($D_d=300$~m): (a) negative-feedback rate and (b) blacklist rate in the two GT rounds after the state change.}\label{fig:single_neg}
\end{figure*}

Up to 180~s, all feedback was positive in both schemes. Negative feedback appeared only after the state change at $T_{stop}=200$~s, as predicted by H1 (\autoref{fig:single_neg}). In the two rounds following the state change, SAFE reduced the negative-feedback rate from 53.8\% to 24.7\% and from 55.6\% to 8.3\%. Similarly, the blacklist rate decreased from 66.7\% to 21.4\% at 240~s and from 45.5\% to 12.5\% at 300~s. Over the entire run, 20 distinct honest vehicles (out of 72 evaluated) were blacklisted in TCEMD, whereas this number was 6 (out of 78) in SAFE. All six vehicles blacklisted in SAFE were also blacklisted in TCEMD; in other words, SAFE introduced no new false positives.

SAFE also improved the decisions of vehicles, although it does not change the decision rule. All rejected warnings occurred after $T_{stop}$, between 206~s and 262~s. With $D_d=300$~m, TCEMD produced 13 rejected warnings and SAFE produced 10 (\autoref{tab:decisions}). The first seven rejections, up to 240~s, were common to both schemes. After the GT round at 240~s, in which TCEMD blacklisted 16 vehicles and SAFE blacklisted 6, TCEMD produced six further rejections, while SAFE produced three. SAFE thus improves the input of the decision by preserving the trust of honest senders.

\begin{table}[!htbp]
\centering
\caption{Rejected warnings in the single-event scenario. In total, 81 vehicles reached $D_d$; in TCEMD, one of them had received no report, leaving 80 decisions.}
\label{tab:decisions}
\begin{tabular}{lcc}
\toprule
 & TCEMD & SAFE \\
\midrule
$D_d = 300$ m & 13 / 80 (16.3\%) & 10 / 81 (12.3\%) \\
$D_d = 200$ m & 5 / 80 (6.3\%) & 5 / 81 (6.2\%) \\
\bottomrule
\end{tabular}
\end{table}

\subsection{Multi-event scenario}
\label{sec:multi}

In this scenario, three events of different types were placed on the vehicle routes (\autoref{tab:multi}). Their witness areas and active periods partly overlap, which allows a vehicle to be evaluated for more than one event in the same round. In this configuration, $D_d - D_w \le 200$~m, which satisfies Eq.~(\ref{eq:constraint}).

\begin{table*}[!htbp]
\centering
\caption{Event parameters of the multi-event scenario (all other parameters as in \autoref{tab:setup})}
\label{tab:multi}
\begin{tabular}{llll}
\toprule
 & Event 1 & Event 2 & Event 3 \\
\midrule
Type & 1 & 2 & 3 \\
$T_{start}$ / duration & 60 s / 100 s & 250 s / 50 s & 150 s / 100 s \\
$T_{stop}$ / $T_{lasting}$ & 160 s / 210 s & 300 s / 325 s & 250 s / 300 s \\
$D_w$ / $D_d$ / $D_i$ & 100 / 200 / 400 m & 400 / 600 / 800 m & 800 / 1000 / 1200 m \\
Location ($x$, $y$) & (1000 m, 400 m) & (3000 m, 400 m) & (2000 m, 400 m) \\
\bottomrule
\end{tabular}
\end{table*}

\begin{figure*}[!htbp]
\centering
\includegraphics[width=\textwidth]{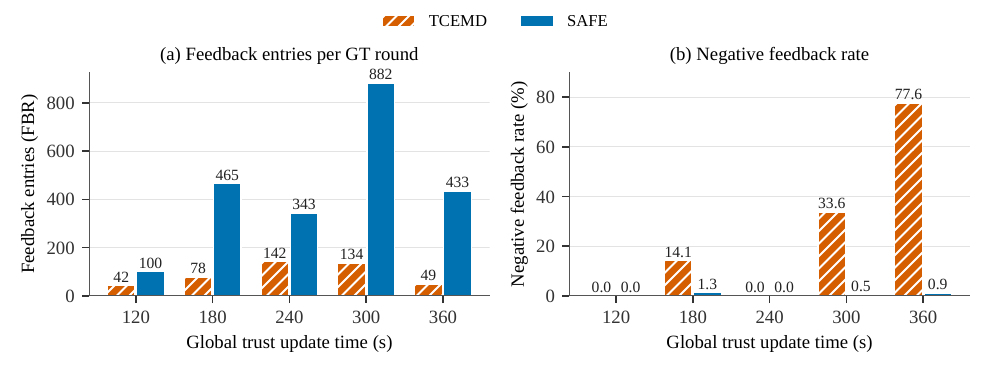}
\caption{Multi-event scenario: (a) feedback entries and (b) negative-feedback rate in each GT round.}\label{fig:multi}
\end{figure*}

\begin{figure*}[!htbp]
\centering
\includegraphics[width=\textwidth]{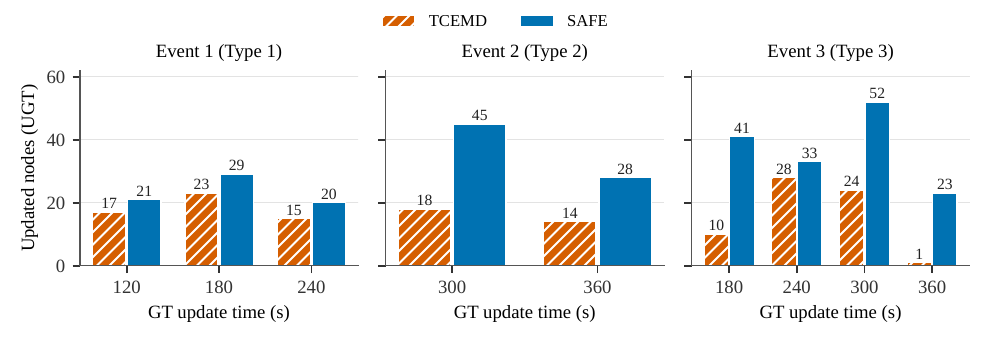}
\caption{Multi-event scenario: vehicles with updated GT per event and GT round.}\label{fig:multi_ugt}
\end{figure*}

SAFE collected 2223 feedback entries against 445 in TCEMD (5.0 times), with the largest difference at 300~s (882 against 134; \autoref{fig:multi}(a)). In TCEMD, the negative-feedback rate reached 14.1\% at 180~s, 33.6\% at 300~s and 77.6\% at 360~s, following the stops of Event 1 (160~s), Event 3 (250~s) and Event 2 (300~s). In SAFE, it remained at or below 1.3\% in every round (\autoref{fig:multi}(b)). Moreover, SAFE updated the GT of more vehicles for every event (\autoref{fig:multi_ugt}). For Event 3, it updated 41 vehicles against 10 at 180~s and 52 against 24 at 300~s.

\autoref{tab:class} presents the node classification. The counts are node-evaluations summed over rounds; for this reason, the totals differ between the schemes, and the shares provide a fairer basis for comparison. In TCEMD, 34 of 150 evaluations (22.7\%) resulted in an untrusted label, whereas in SAFE, only 1 of 292 (0.3\%) did. The TCEMD log counts a blacklisted vehicle as untrusted in every event in which it is evaluated afterwards. This behaviour is evident at 180~s, where six untrusted labels appear for Event 3, although Event 3 received only positive feedback in that round. When the same counting rule is applied to SAFE, its single blacklisted vehicle yields 5 untrusted labels, corresponding to 1.7\% of evaluations, which is still far below the 22.7\% of TCEMD. In terms of distinct vehicles, SAFE blacklisted a single vehicle, whereas TCEMD blacklisted between 11 and 34.

\begin{table*}[!htbp]
\centering
\caption{Node classification in the multi-event scenario, counted as node-evaluations over all GT rounds (Eq.~(\ref{eq:label}))}
\label{tab:class}
\begin{tabular}{llrrrr}
\toprule
Event & Scheme & Untrusted & Suspicious & Honest & Total \\
\midrule
1 & TCEMD & 6 & 1 & 48 & 55 \\
  & SAFE  & 1 & 1 & 68 & 70 \\
\midrule
2 & TCEMD & 11 & 0 & 21 & 32 \\
  & SAFE  & 0 & 2 & 71 & 73 \\
\midrule
3 & TCEMD & 17 & 2 & 44 & 63 \\
  & SAFE  & 0 & 0 & 149 & 149 \\
\midrule
All & TCEMD & 34 (22.7\%) & 3 (2.0\%) & 113 (75.3\%) & 150 \\
    & SAFE  & 1 (0.3\%) & 3 (1.0\%) & 288 (98.6\%) & 292 \\
\bottomrule
\end{tabular}
\end{table*}

\begin{figure*}[!htbp]
\centering
\includegraphics[width=\textwidth]{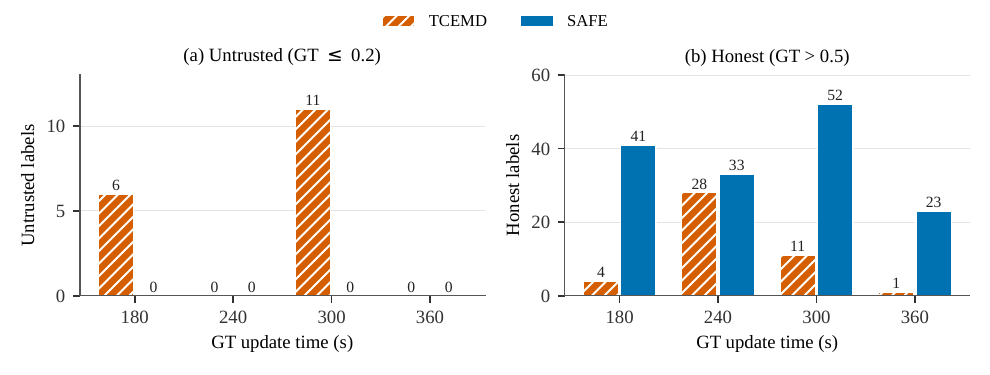}
\caption{Event 3 in the multi-event scenario: (a) untrusted and (b) honest labels per GT round.}\label{fig:event3}
\end{figure*}

\autoref{fig:event3} follows Event 3, which has the widest witness area, over time. TCEMD produced 6 and 11 untrusted labels at 180~s and 300~s, respectively, and at most 28 honest labels per round. In contrast, SAFE produced no untrusted label for this event and up to 52 honest labels per round.

\subsection{Contribution of the SAFE elements}
\label{sec:mech}

The feedback logs of the single-event scenario ($D_d=300$~m) contain a time stamp for every feedback entry. This allows the effect of SAFE to be decomposed as follows:
\begin{itemize}
    \setlength{\itemsep}{0pt}
    \item \emph{Recording between $D_d$ and $D_w$ (C1).} The number of entries created at the first evaluation of each reporter increased from 187 in TCEMD to 304 in SAFE. Considering only these entries, the negative-feedback rate of SAFE was 35.5\% at 240~s and 14.6\% at 300~s, compared with 53.8\% and 55.6\% in TCEMD.
    \item \emph{Evidence collected inside the witness area (C1 and C2).} SAFE produced 178 additional entries after the first evaluation of each reporter, only 3 of which were negative. These entries reach the CDU through the exit report (C2) and further reduced the negative-feedback rate to 24.7\% and 8.3\%. Overall, each reporter produced 9.8 entries in SAFE against 4.0 in TCEMD, and at 240~s the CDU had 5.2 entries per updated vehicle against 2.7.
    \item \emph{Re-evaluation of the same sender.} Only 14 of 468 reporter--target pairs were evaluated twice, and none of them changed its score. Hence, the benefit of SAFE stems from new and more recent evidence rather than from the revision of earlier evaluations.
\end{itemize}
In the single-event scenario, the number of negative entries remained similar (45 in TCEMD and 42 in SAFE), while the number of positive entries increased substantially. The improvement is therefore attributable to the additional and more recent positive evidence, which outweighed the remaining stale entries in Eq.~(\ref{eq:gt}). In the multi-event scenario, where the witness areas are wider, the number of negative entries itself decreased from 94 to 14. Both elements of SAFE contribute to these results. Recording up to the witness area improves the first report, and the evidence collected inside the witness area, delivered by the exit report, improves it further. Since this in-area evidence reaches the CDU through the exit report, C1 and C2 operate as complementary parts of a single mechanism.

\subsection{Effect of the decision distance}
\label{sec:dd}

\begin{figure*}[!htbp]
\centering
\includegraphics[width=\textwidth]{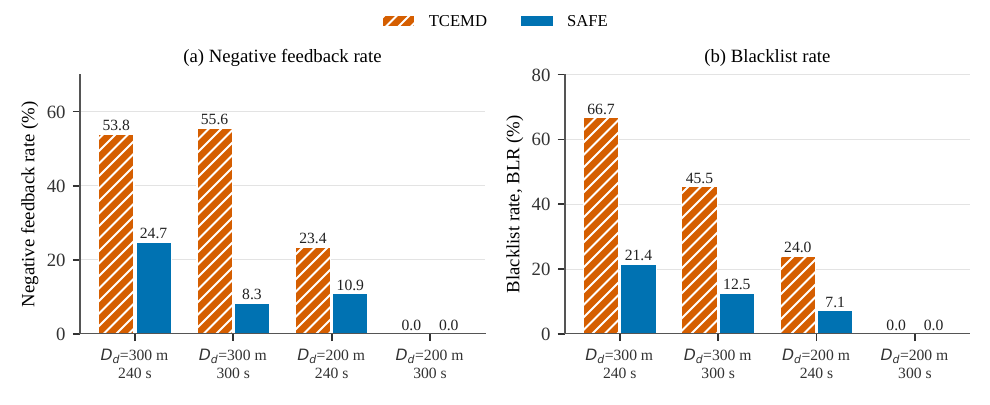}
\caption{Effect of the decision distance in the single-event scenario: (a) negative-feedback rate and (b) blacklist rate for $D_d=300$~m and $D_d=200$~m in the two GT rounds after the state change.}\label{fig:dd}
\end{figure*}

The single-event scenario was repeated with $D_d = 200$~m for both schemes. A shorter decision distance reduced errors in both schemes (\autoref{fig:dd}). In TCEMD, the negative-feedback rate in the two rounds after the state change decreased from 53.8\% and 55.6\% to 23.4\% and 0\%, and the blacklist rate from 66.7\% and 45.5\% to 24.0\% and 0\%. At $D_d=200$~m, SAFE further reduced the negative-feedback rate from 23.4\% to 10.9\% and the blacklist rate from 24.0\% to 7.1\% at 240~s, and both schemes produced no errors at 300~s. TCEMD blacklisted six distinct vehicles and SAFE blacklisted two, which were again among those of TCEMD. Rejected warnings decreased to five in both schemes (\autoref{tab:decisions}). The feedback volume changed only slightly with $D_d$, from 187 to 213 entries in TCEMD and from 482 to 496 in SAFE.

These results confirm the stale-witness mechanism. With a smaller $D_d - D_w$, a TCEMD vehicle stops recording closer to the witness area; as a result, its evidence is more recent. This corresponds to the effect that SAFE achieves by recording for a longer period, which explains why the absolute gain of SAFE is smaller at $D_d=200$~m. In the multi-event scenario, increasing $D_d$ by 100~m for all three events (SAFE only, with $D_d - D_w$ up to 300~m) left the performance of SAFE almost unchanged: the negative-feedback rate remained at or below 1.8\%, and two untrusted labels were produced instead of one. Moreover, the multi-event configuration achieved low error rates with $D_d/D_w = 1.25$ and $1.5$. These findings indicate that the gap $D_d - D_w$, rather than the ratio $D_d/D_w$, is the relevant design quantity. This gap must satisfy Eq.~(\ref{eq:constraint}) and leave sufficient reaction time.

\subsection{Overhead}
\label{sec:overhead}

\textbf{Communication.} The cost of SAFE is additional feedback traffic from vehicles to RSUs (\autoref{tab:overhead}). By design, the number of feedback messages per vehicle and event increases from one to at most two, owing to the exit report. The number of feedback entries increases by 2.3--2.6 times in the single-event scenarios and 5.0 times in the multi-event scenario, as each report covers more senders. EM broadcasts remain unchanged, given that SAFE does not alter when or how often witnesses broadcast. Each entry contains only identifiers, a score and a time stamp, and FBs are sent only when a vehicle witnesses or leaves an event.

\begin{table}[!htbp]
\centering
\caption{Feedback volume (entries received by the CDU over the whole run)}
\label{tab:overhead}
\begin{tabular}{lrrr}
\toprule
Scenario & TCEMD & SAFE & Ratio \\
\midrule
Single event, $D_d=300$ m & 187 & 482 & 2.6 \\
Single event, $D_d=200$ m & 213 & 496 & 2.3 \\
Multi event & 445 & 2223 & 5.0 \\
\bottomrule
\end{tabular}
\end{table}

\textbf{Storage and computation on the OBU.} As in TCEMD, a vehicle stores at most one record (sender, state, time stamp) per sender and event. SAFE keeps these records for a longer period, from the decision point to the exit from the witness area, while still keeping a single record per sender. Accordingly, storage is bounded by the number of distinct senders heard per event, which was about ten per reporter in the single-event scenario. Building a report requires a linear pass over these records, $O(|Rec_e|)$, performed at most twice per event.

\textbf{Decision latency.} The decision is made at $D_d$ with the same rule and the same inputs available at that moment. All additional processing takes place after the decision; hence, SAFE introduces no delay into the safety-relevant action.

\textbf{CDU.} The CDU processes 2.3--5.0 times more entries per round. The computation in Eq.~(\ref{eq:gt}) is linear in the number of entries, and the CDU is equipped with sufficient computational resources \cite{TCEMD,AATMS}.

\section{Discussion}
\label{sec:discussion}

This section evaluates the hypotheses in the light of the results, discusses the security implications of SAFE, and summarises the limitations of the study together with future directions.

\subsection{Evaluation of the hypotheses}

The results support H1. In all simulated scenarios, negative feedback appeared only after an event changed state. In TCEMD, this feedback led to the blacklisting of up to 66.7\% of the updated vehicles in the single-event scenario and 78.6\% (11 of 14) in the Event~2 evaluation at 360~s in the multi-event scenario.

The results also support H2. SAFE reduced the negative-feedback rate, the blacklist rate and node misclassification in all scenarios, at the cost of 2.3--5.0 times more feedback entries. In the single-event scenario, the improvement resulted mainly from additional and more recent positive evidence that outweighed the stale entries. In the multi-event scenario, the stale entries themselves were also largely eliminated. Both elements of SAFE contributed: recording up to the witness area improved the first report, and the evidence delivered by the exit report improved it further.

\textbf{Application.} SAFE directly benefits honest vehicles, whose warnings remain trusted, and the CDU, whose blacklist contains fewer false positives. A false blacklisting excludes an honest warning source from all subsequent events. By preventing such errors, SAFE preserves the dissemination coverage on which event-based warnings depend. This is particularly important on highways, where an incident affects a long upstream section of traffic.

\subsection{Security considerations}
\label{sec:security}

SAFE is designed for the feedback-collection layer and can be combined with the attack-detection mechanisms of existing trust schemes. Its interaction with adversarial behaviour can be characterised as follows:
\begin{itemize}
    \setlength{\itemsep}{0pt}
    \item \emph{Effect of the exit report.} SAFE allows each vehicle to send one additional feedback report per event. As the CDU retains only the latest entry for each $\langle$reporter, target, event$\rangle$ triple (\autoref{sec:tcemd}), a repeated evaluation replaces the earlier one and does not add weight. The longer recording window enables a vehicle to evaluate more targets; Eq.~(\ref{eq:gt}) already weights each entry by the trust of its reporter, which limits the influence of low-trust reporters. Bounding the number of targets per report provides an additional safeguard against colluding vehicles.
    \item \emph{Honest-majority requirement.} Each feedback entry is judged against the observation of the evaluator, and GT aggregates the entries of many evaluators. Correct operation therefore requires honest evaluators to outnumber colluding ones among the witnesses of an event. When all witnesses are malicious, an out-of-band source, such as infrastructure sensors, is required to establish the true event state. Determining the minimum fraction of honest witnesses is a direct extension of this study.
    \item \emph{Incentive structure.} Correct reporting is rewarded through a higher GT and, consequently, greater influence on decisions and trust computation. By protecting honest vehicles from stale-feedback penalties, SAFE strengthens this reward.
    \item \emph{Trust in infrastructure.} The trusted RSU and CDU assumption can be relaxed by replicating the CDU and requiring agreement among replicas, by cross-checking the reports forwarded by different RSUs, or by storing feedback on a distributed ledger, as in DID- and ledger-based reputation systems \cite{DIVA}. Being independent of where GT is computed, SAFE is compatible with each of these architectures.
\end{itemize}

\subsection{Limitations and future directions}
\label{sec:limits}

The evaluation was designed to isolate the stale-witness problem and to measure the effect of SAFE on it. This focus entails the following limitations, each of which defines a direction for future work:
\begin{itemize}
    \setlength{\itemsep}{0pt}
    \item \emph{Number of runs.} Each configuration was simulated as a single deterministic run. Repeating the scenarios over multiple traffic realisations will quantify the variability of the results.
    \item \emph{Baseline scheme.} SAFE was compared with TCEMD, the distance-based scheme on which it is built, which isolates the effect of SAFE from that of the trust formula. Integrating SAFE into other distance-based schemes, such as HTEMD \cite{HTEMD}, will demonstrate its generality.
    \item \emph{Component analysis.} The contributions of the SAFE elements were decomposed using the time-stamped simulation logs (\autoref{sec:mech}). A controlled analysis in which each element is enabled separately will complement this decomposition.
    \item \emph{Adversarial scenarios.} The evaluation focused on fault tolerance in a network of honest vehicles. Testing SAFE under recommendation, false-data and collusion attacks will show how its fault-tolerance gains combine with attack detection.
    \item \emph{Scenario diversity.} The scenarios covered a highway with up to three events of different types and two decision distances. Urban topologies and additional decision distances will broaden the analysis.
\end{itemize}

\section{Conclusion}
\label{sec:conclusion}

This study defined the stale-witness problem in event-based VANET trust management: honest vehicles are penalised when their correct but outdated messages are evaluated against a new event state. To address this problem, SAFE was proposed as a feedback-collection mechanism in which vehicles continue to record event messages after their decision and throughout the witness area, and send an updated feedback report when they leave it. SAFE was compared with TCEMD in single-event, multi-event and decision-distance scenarios. The main findings are as follows:
\begin{itemize}
    \setlength{\itemsep}{0pt}
    \item In the single-event scenario, SAFE reduced the negative-feedback rate after the state change from 53.8--55.6\% to 8.3--24.7\% and the number of distinct blacklisted honest vehicles from 20 to 6.
    \item In the multi-event scenario, the negative-feedback rate remained at or below 1.3\% in SAFE, compared with up to 77.6\% in TCEMD, and the share of evaluations ending in an untrusted label decreased from 22.7\% to at most 1.7\%.
    \item SAFE also reduced the number of rejected warnings from 13 to 10 at $D_d=300$~m, thereby improving the decisions of vehicles without changing the decision rule.
    \item These gains were obtained with 2.3--5.0 times more feedback entries and without any delay in the decision.
    \item A shorter decision distance reduced stale feedback in both schemes, and the gap $D_d - D_w$, bounded by the communication range, was identified as the relevant design quantity.
\end{itemize}
These findings confirm that keeping feedback evidence up to date constitutes an effective means of protecting honest vehicles in event-based trust systems. Future work will extend the evaluation to multiple traffic realisations, integrate SAFE into other distance-based schemes and evaluate its performance under recommendation, false-data and collusion attacks.

\section*{Acknowledgments}
This study was supported by the Scientific and Technological Research Council of Türkiye (TÜBİTAK) under Grant Number 124E017. The author thanks TÜBİTAK for its support.

\section*{Declarations}
\textbf{Conflict of interest:} The author declares that there is no conflict of interest regarding the publication of this paper.

\noindent\textbf{Data availability:} The data supporting the findings of this study are available from the author upon reasonable request.

\noindent\textbf{Use of artificial intelligence:} During the preparation of this manuscript, the author used Claude (Anthropic) for language refinement, proofreading and figure generation from the simulation logs. The author reviewed and edited all AI-assisted content and takes full responsibility for the content of the article.

\end{document}